\documentclass[11pt]{article}
\usepackage{fullpage}
\usepackage{psfrag}
\usepackage{amssymb}
\usepackage{amsmath}
\usepackage{amsfonts}
\usepackage{epsfig} 
\usepackage{comment} 
\usepackage{float}     
\restylefloat{figure}  

\newcommand{\ba}{\begin{array}}
\newcommand{\ea}{\end{array}}
\newcommand{\bi}{\begin{itemize}}
\newcommand{\ei}{\end{itemize}}
\newcommand{\bcm}{\left[ \begin{array}}
\newcommand{\ecm}{\end{array} \right]}
\newcommand{\ben}{\begin{enumerate}}
\newcommand{\een}{\end{enumerate}}
\newcommand{\bea}{\begin{eqnarray}}
\newcommand{\eea}{\end{eqnarray}}
\newcommand{\bas}{\begin{eqnarray*}}
\newcommand{\eas}{\end{eqnarray*}}
\newcommand{\beqs}{\begin{equation*}}
\newcommand{\eeqs}{\end{equation*}}
\newcommand{\beq}{\begin{equation}}
\newcommand{\eeq}{\end{equation}}
\newcommand{\bml}{\begin{multline}}
\newcommand{\eml}{\end{multline}}
\newcommand{\bs}{\begin{slide}}
\newcommand{\es}{\end{slide}}
\newcommand{\bc}{\begin{center}}
\newcommand{\ec}{\end{center}}
\newcommand{\bt}{\begin{table}}
\newcommand{\et}{\end{table}}

\newcommand{\sn}{\mbox{\,sn\,}}
\newcommand{\cn}{\mbox{\,cn\,}}
\newcommand{\dn}{\mbox{\,dn\,}}
\newcommand{\eps}{\epsilon}

\title{Instabilities of one-dimensional stationary solutions of the cubic nonlinear Schr\"{o}dinger equation}
\date{\today}
\author{Roger J. Thelwell\dag, John D. Carter\ddag, Bernard Deconinck\dag \\ \\ 
\begin{small}
\begin{tabular}{ll}
  \dag Department of Applied Mathematics& \ddag Mathematics Department \\
  \, University of Washington & \,  Seattle University\\
  \, Seattle, WA 98195-2420 & \, Seattle, WA 98122
\end{tabular}
\end{small}}

\begin{document}

\maketitle

\begin{abstract}
\label{s:abstract}

The two-dimensional cubic nonlinear Schr\"{o}dinger equation
admits a large family of one-dimensional bounded traveling-wave 
solutions.  All such solutions may be written in terms of an 
amplitude and a phase.  Solutions with piecewise constant phase have been 
well studied previously. Some of these solutions 
were found to be stable with respect to one-dimensional perturbations. 
No such solutions are stable with respect to two-dimensional perturbations.
Here we consider stability of the larger class of solutions whose phase is 
dependent on the spatial dimension of the one-dimensional wave form.
We study the spectral stability of such nontrivial-phase solutions numerically, 
using Hill's method. We present evidence which 
suggests that all such nontrivial-phase solutions are 
unstable with respect to both one- and two-dimensional perturbations.  
Instability occurs in all cases: for both the elliptic and hyperbolic 
nonlinear Schr\"{o}dinger equations, and in the focusing and defocusing case.  
\end{abstract}

\section{Introduction}
\label{s:intro}

The cubic nonlinear Schr\"{o}dinger (NLS) equation in two spatial dimensions is
given by
\begin{equation}
\label{E:NLS}
i\psi_t + \alpha \psi_{xx} + \beta \psi_{yy} + |\psi|^2 \psi = 0.
\end{equation}
The NLS equation is said be {\em focusing} or
{\em attractive} in the $x$-dimension if $\alpha >0$.  If $\alpha < 0$,
the NLS equation is said to be {\em defocusing} or {\em repulsive} in the 
$x$-dimension.  Similarly, the sign of $\beta$ leads to focusing 
or defocusing in the $y$-dimension.
The NLS equation is called {\em hyperbolic} if 
$\alpha \beta <0$ and {\em elliptic} if $\alpha \beta > 0$. 

Equation \eqref{E:NLS} admits a large family of one-dimensional bounded traveling-wave solutions.
All such solutions, up to Lie group symmetries~\cite{sulemsulem}, may be written in the form~\cite{carr1,carr2}
\begin{equation}
\label{E:soln_form}
\psi(x,t)=\phi(x) e^{i\theta(x) + i \lambda t},
\end{equation}
where $\phi(x)$ and $\theta(x)$ are real-valued functions, and $\lambda$ is a
real constant.  Solutions of the form \eqref{E:soln_form} are possible if
\begin{subequations}
\label{E:defs}
\begin{eqnarray}
\phi^2(x)&=& \displaystyle{\alpha \left(-2 k^2 \sn^2(x,k)+B \right),}\label{E:phi_def} \\
\theta(x)&=& \displaystyle{c\int_0^x\phi^{-2}(\xi)d\xi,}  \label{E:theta_def} \\
\lambda&=& \displaystyle{\frac{1}{2}\alpha (3B-2-2k^2),} \label{E:omega_def} \\
c^2&=& \displaystyle{-\frac{\alpha^2}{2} B(B-2k^2)(B-2),} \label{E:c_def}
\end{eqnarray}
\end{subequations}
where $c$ is a real constant.
Here $k \in [0,1]$ is the elliptic modulus of the 
Jacobi elliptic sine function, $\sn(x,k)$.   
The function $\sn(x,k)$ is periodic if $k\in[0,1)$, with 
period given by $L = 4K$, where $K=K(k)$ is defined by 
\begin{equation}
K(k)=\int_{0}^{\pi/2}\left(1-k^2\sin^2 x\right)^{-1/2} \, dx,
\end{equation}
and is known as the complete elliptic integral of the first kind.
When $k=0$, $\sn(x,0) = \sin(x)$ with $L = 2 \pi$.  As 
$k$ approaches 1, $\sn(x,k)$ approaches $\tanh(x)$ and $L$ approaches 
infinity~\cite{byrd}. Although $\phi(x)$ inherits the periodicity of $\sn(x,k)$, the solution 
$\psi(x,t)$ is
typically {\em not} $L$-periodic in the $x$-dimension, because the periods 
of $e^{i\theta}$ and $\phi$ are typically non-commensurate.

The solution $\psi$ is said to have 
{\em trivial-phase} (TP) if $\theta(x)$ is (piecewise) constant and  
{\em nontrivial-phase} (NTP) if $\theta(x)$ is not constant. Equivalently, the solution $\psi$ has TP if $c=0$, and has NTP if $c \ne 0.$
For every choice of $\alpha$ and $\beta$,~\eqref{E:defs} specifies a 
two-parameter family of NLS solutions with the free parameters $k$ and $B$.  
Without loss of generality, we choose both $\alpha$ and $\beta$ to be 
$\pm 1$.  The phase contribution $\theta(x)$ given in \eqref{E:theta_def} 
implicitly depends on $\alpha$ and $B$ in both \eqref{E:phi_def} 
and \eqref{E:c_def}. In order for $\phi$ and $\theta$ to be real-valued 
functions, we need $B \in  [2 k^2 , 2]$ if $\alpha =1$ or
$B \le 0$ if $\alpha = -1$. 
\begin{figure}
 \psfrag{Stokes}{Stokes'}
 \psfrag{sn}{sn}
 \psfrag{cn}{cn}
 \psfrag{dn}{dn}
 \psfrag{soliton}{soliton}
 \psfrag{grey}{gray}
 \psfrag{dark}{dark}
 \psfrag{k}{k}
 \psfrag{B}{B}
 \psfrag{0}{0}
 \psfrag{-1}{-1}
 \psfrag{2}{2}
 \psfrag{1}{1}
 \bc
  \begin{tabular}{cc}
   \epsfig{file=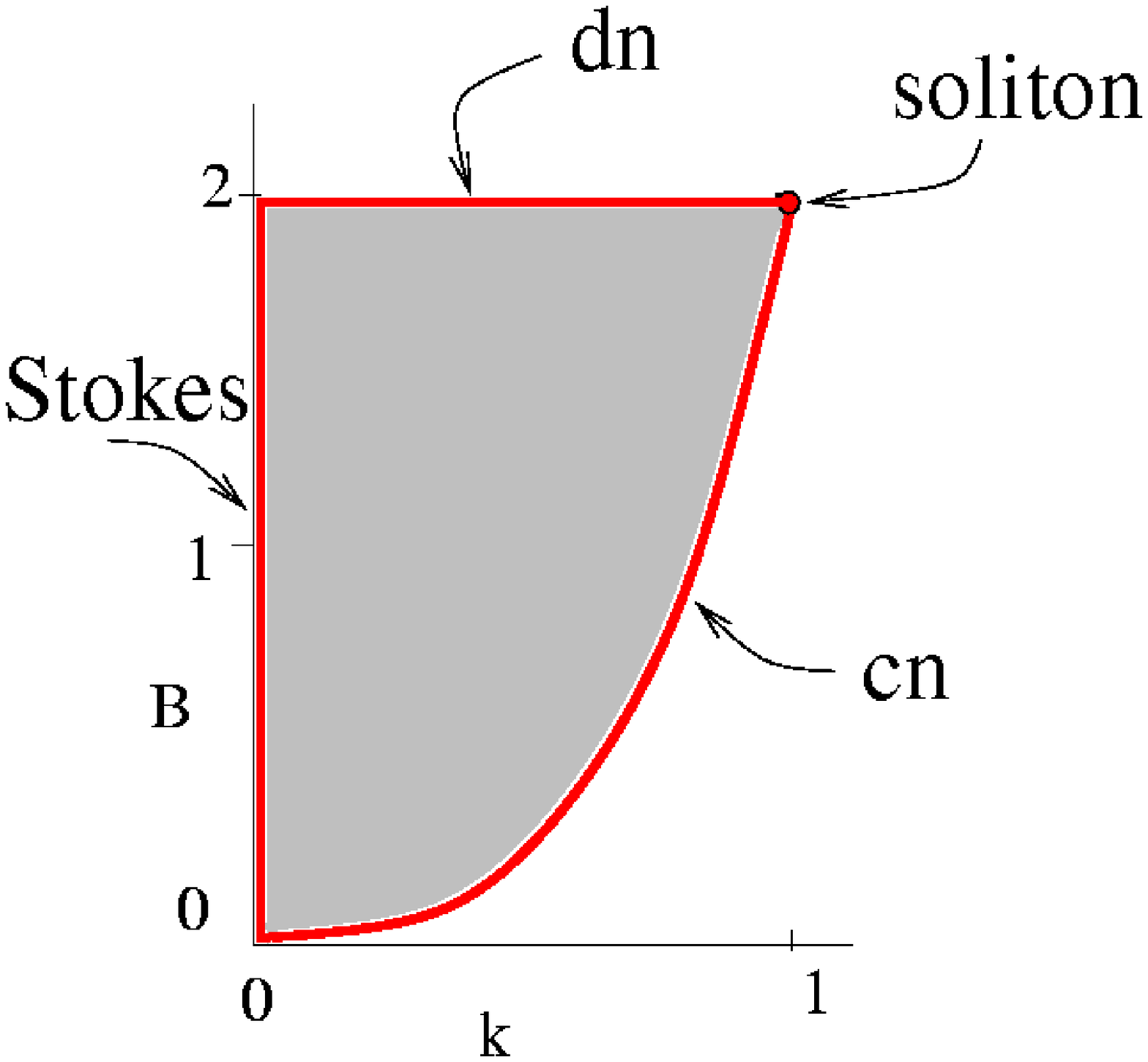, width = 7 cm} &
   \epsfig{file=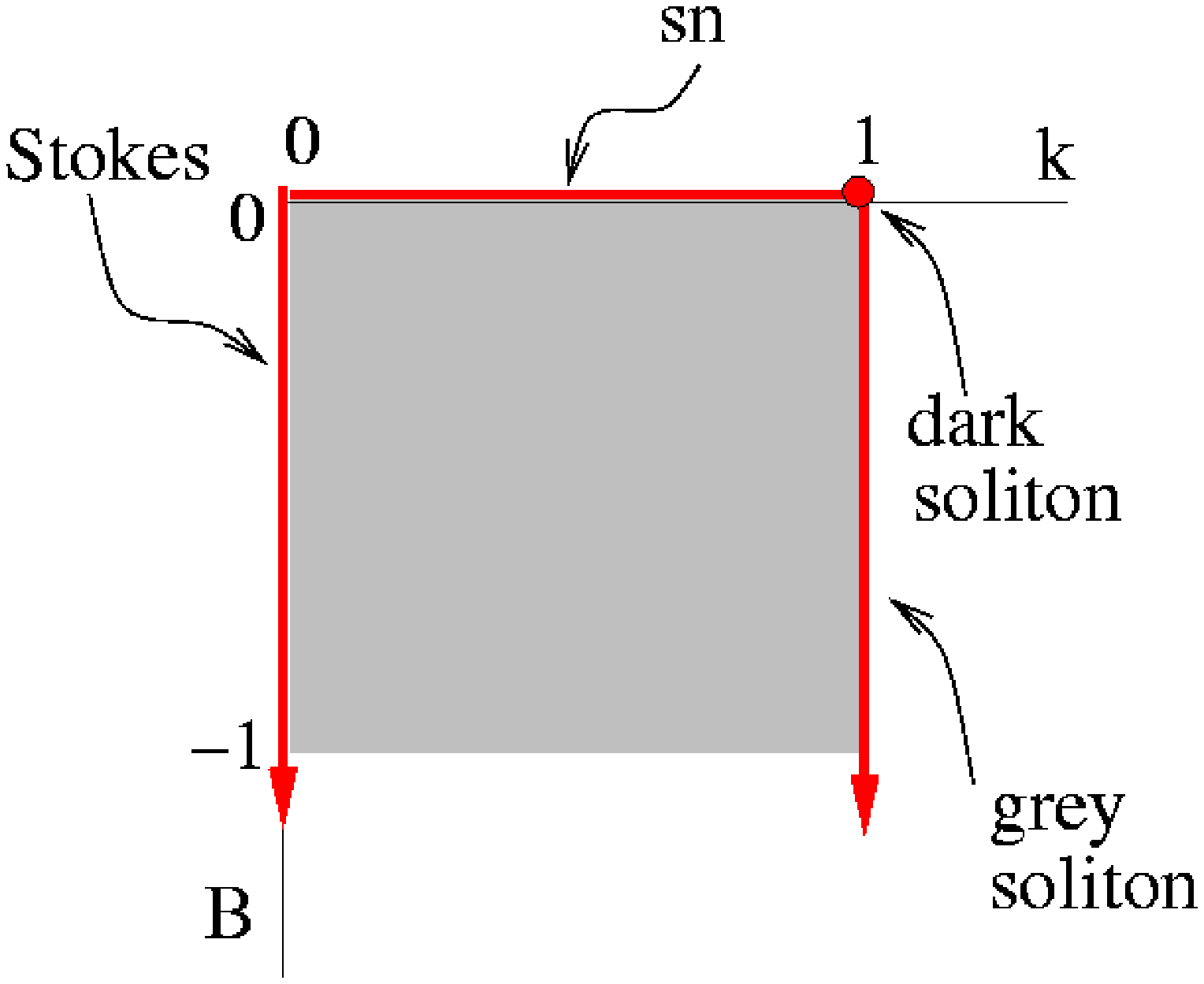, width = 8 cm} \\
   (a) $\alpha = +1$ &  (b) $\alpha = -1$
  \end{tabular}
  \caption{Admissible parameter space for solutions of the form given in \eqref{E:soln_form} for (a) focusing and 
(b) defocusing regimes.  The shaded interior region corresponds to NTP solutions. The region in the defocusing regime 
is unbounded below.\label{F:parameter_space}}
 \ec
\end{figure}
Figure~\ref{F:parameter_space} displays the regions of $(k,B)$-parameter
space that correspond to NTP solutions of the elliptic and hyperbolic 
NLS equations.  By varying $B$ so that $c \to 0$, $\theta$ approaches a 
(piecewise) constant, and the NTP solutions reduce to 
one of five types of TP solutions: 
({\em i}) a Stokes' plane wave, ({\em ii}) a cn-type solution,
({\em iii}) a dn-type solution, ({\em iv}) an sn-type solution,   
({\em v}) a soliton-type solution.
The limiting solutions correspond to the boundaries of the regions in 
Fig.~\ref{F:parameter_space}.  
Table~\ref{T:tp_limit} provides the values of $k$ and $B$ that 
cause~\eqref{E:soln_form} to reduce to TP solutions, and also 
gives the explicit expression for $\psi$ in the TP limit. 
Gray solitons, all of which are NTP solutions, result 
when $k=1$ and $B<0$.
An overview of stationary NLS solutions is given in~\cite{carr1,carr2}.  
Details of the Jacobi elliptic functions cn, dn and sn may be found 
in~\cite{byrd}. 

\begin{table}[t]
\bc
\begin{tabular}{|c|ll|c|l|}
\hline
 & $k$ value & $B$ value & Solution type & \hspace*{1cm} $\psi$  \\
\hline
\hline
$\alpha = 1$     & $k = 0$ &  $B\in[2k^2,2]$ & Stokes' plane wave & $\sqrt{B} \, e^{i \lambda t}$\\
     &  $k \in(0,1) $ &  $B=2k^2$ & cn-type & $\sqrt{2} \, k \cn(x,k) \, e^{i(2k^2 - 1)t}$ \\
     &  $k \in(0,1) $ &  $B=2$ & dn-type & $\sqrt{2} \, \dn(x,k) \, e^{i(2-k^2)t}$ \\
&  $k = 1$ &  $B = 2$ & bright soliton & $ \sqrt{2} \, \, \mbox{sech}(x) e^{it}$ \\
\hline
\hline
$\alpha = -1$ & $k = 0$ &  $B\le 0$ & Stokes' plane wave & $\sqrt{B} \, e^{i\lambda t}$\\
     & $k \in(0,1) $ &  $B=0$ & sn-type & $\sqrt{2} \, k \sn(x,k) e^{i(1+k^2)t}$\\
& $k = 1$ &  $B = 0$ & dark soliton & $\sqrt{2} \, \tanh(x) \, e^{i2t}$ \\
\hline
\end{tabular}
\caption{ Parameter values $(k,B)$ which reduce NTP solutions to TP solutions.
\label{T:tp_limit}}
\ec
\end{table}

The stability analysis of TP solutions is well investigated,
see for example~\cite{zr,MYS,KuzRubZak,FilFraYun,rr2,JCBDtp}. 
While some TP solutions 
are stable under one-dimensional perturbations (the bright soliton~\cite{kivpel,JCBDsoliton} and 
sn-type solutions~\cite{JCBDtp}), all TP solutions are known to be unstable under two-dimensional perturbations~\cite{JCBDtp}.
We know of only the 1981 work of Infeld and 
Ziemkiewicz~\cite{infeldz} for results regarding the stability of some NTP 
solutions of the NLS equation.  An additional damping term may lead 
to stable NTP solutions.  However, without this term they found that all NTP solutions 
they considered are unstable. Our results agree with their conclusions. The work presented here differs
from~\cite{infeldz} in that we consider the entire parameter space of NTP solutions, so that all one-dimensional
stationary solutions of the NLS equation have now been investigated.

The $(k,B)$-parameter space of the NTP solution is two-dimensional, 
whereas the $(k,B)$-parameter space of TP solutions
is essentially one-dimensional; it forms the boundary of the NTP 
$(k,B)$-parameter spaces shown in Fig.~\ref{F:parameter_space}. Thus
the TP solutions are only a co-dimension one subset of all bounded traveling-wave 
solutions of \eqref{E:NLS}.  The aim of this work is to investigate 
if \eqref{E:NLS} has {\em any} one-dimensional traveling-wave solutions that
are stable with respect to either one- or two-dimensional perturbations.  Since
no solutions are known that are stable with respect to two-dimensional perturbations 
in the TP setting, we focus especially on such perturbations, although
one-dimensional perturbations are also considered as a special case.

We investigate the spectral stability
of all NTP solutions~\eqref{E:soln_form}, for all choices of $\alpha=\pm1$ and $\beta=\pm1$.
Although all NTP solutions are found to be unstable, our investigations do produce important 
information about the nature of the instabilities of these NTP solutions.

\section{The linear stability problem}
\label{S:linear}

In order to study the linear stability of NTP solutions
of the NLS equation, we consider perturbed solutions of the form
\begin{equation}
\psi_{_p}(x,y,t)=(\phi(x)+\epsilon u(x,y,t)+i \epsilon
v(x,y,t) + \mathcal{O}(\eps^2)) \, e^{i\theta(x) + i\lambda t},
\label{E:npert}
\end{equation}
where $u(x,y,t)$ and $v(x,y,t)$ are real-valued functions, $\epsilon$ is a
small real parameter and $\phi(x) \, e^{i\theta(x)+i\lambda t}$ is
an NTP solution of NLS.  
Substituting (\ref{E:npert})
in (\ref{E:NLS}), linearizing and separating real and imaginary
parts leads to
\begin{subequations}
\begin{equation}
\lambda u-3\gamma\phi^2u-\beta
u_{yy}+ \alpha c^2 \frac{1}{\phi^4}u-2\alpha c\frac{1}{\phi^{3}}\phi_x v+2\alpha
c\frac{1}{\phi^{2}}v_x-\alpha u_{xx}=-v_t,\label{nreale}
\end{equation}
\begin{equation}
\lambda v-\gamma\phi^2v-\beta v_{yy}+\alpha c^2\frac{1}{\phi^{4}}v+2\alpha
c\frac{1}{\phi^{3}}\phi_x u-2\alpha c\frac{1}{\phi^{2}}u_x-\alpha v_{xx}=u_t.\label{nimage}
\end{equation}
\label{nevaluepde}
\end{subequations}
Since \eqref{nevaluepde} does not 
depend on $y$ or $t$ explicitly, we may assume that $u(x,y,t)$ and 
$v(x,y,t)$ have the forms 
\begin{subequations}
\begin{equation}
u(x,y,t)=U(x,\rho,\Omega) \, e^{i\rho y+\Omega t}+c.c.,
\end{equation}
\begin{equation}
v(x,y,t)=V(x,\rho,\Omega) \, e^{i\rho y+\Omega t}+c.c.,
\end{equation}
\label{uvnontrivial}
\end{subequations}
where $\rho$ is a real constant, $U(x)$ and $V(x)$ are complex-valued
functions,  $\Omega$ is a complex constant and $c.c.$ denotes complex
conjugate.  Notice that $\rho$ is the transverse wavenumber of the 
perturbation and $\Omega$ is the exponential growth rate associated 
with $\rho$. 
If bounded $U,V$ exist such that $\Omega$ has a positive real part, 
then the amplitudes of the perturbations grow
exponentially in time and the unperturbed solution is unstable.

Upon substitution, (\ref{nevaluepde}) yields the spectral problem 
\begin{subequations}
\label{E:neval}
\begin{equation}
\lambda U-3\gamma\phi^2U+\beta\rho^2
U+\alpha c^2\frac{1}{\phi^{4}}U-2\alpha c\frac{1}{\phi^{3}}\phi_xV+2\alpha
c\frac{1}{\phi^{2}}V_x-\alpha U_{xx}=-\Omega V,
\end{equation}
\begin{equation}
\lambda V-\gamma\phi^2V+\beta\rho^2 V+\alpha c^2\frac{1}{\phi^{4}}V+2\alpha
c\frac{1}{\phi^{3}}\phi_x U-2\alpha c\frac{1}{\phi^{2}}U_x-\alpha
V_{xx}=\Omega U.
\end{equation}
\end{subequations}
If $c=0$, then \eqref{E:neval} reduces 
to the stability problem for TP solutions. 
This case is
examined in~\cite{MYS,infeldz,alesh,kart,JCHStp,JCBDtp} and others. 
Using the linear system \eqref{E:neval}, we are now able to 
investigate the stability of all NTP solution numerically.  
We only consider the stability of NTP solutions in this paper.
The reader may wish to consult some of the above mentioned references
for the stability analysis of their limiting special cases.

\section{The numerical investigation of spectral stability: Hill's method}
\label{s:numerics}

The main difficulty for the numerical investigation of \eqref{E:neval}
is the size of the parameter space involved.  For every choice of the 
equations parameters $\alpha,
\beta$ and solution parameter pairs $(k,B)$, the spectrum of \eqref{E:neval}
needs to be computed for a range of $\rho$ values in order to determine stability or to
analyze any instabilities.  An efficient numerical 
method is necessary.  Hill's method
allows for the systematic and efficient exploration of the 
large phase space encountered here, due to its exponential convergence~\cite{BDNKhill}.

To apply Hill's method, Fourier expansions are needed for all coefficient 
functions of \eqref{E:neval}.  Using the complex Fourier form, we have
\begin{equation}
\label{E:phihat}
\begin{array}{rr}
\displaystyle{\phi^2(x) = \sum_{n = -\infty}^{\infty} Q_n \, e^{i 2 n \pi x / L},} &  \qquad
\displaystyle{\phi^{-2}(x) = \sum_{n = -\infty}^{\infty} R_n \, e^{i 2 n \pi x / L},} \\
\displaystyle{\phi^{-4}(x) = \sum_{n = -\infty}^{\infty} S_n \, e^{i 2 n \pi x / L},} & \qquad 
\displaystyle{\phi^{-3}(x) \phi'(x) = \sum_{n = -\infty}^{\infty} T_n \, e^{i 2 n \pi x / L}}, 
\end{array}
\end{equation}
where $Q_n,R_n,S_n$ and $T_n$ are Fourier coefficients. 
Note that $\phi^2(x)$ has period $L/2$ and that $\phi(x)$ is never zero except in some TP limit cases. 

The periodicity of the coefficient functions in \eqref{E:neval} allows us 
to decompose the eigenfunction components $U$ and $V$ of~\eqref{E:neval} in a 
Fourier-Floquet form
\beq
\label{E:UVhat}
U(x) = e^{i \mu x} \sum_{n=-\infty}^{\infty} U_n \, e^{i n \pi x/ L}   
\mbox{\quad and \quad}
V(x) = e^{i \mu x} \sum_{n=-\infty}^{\infty} V_n \, e^{i n \pi x/ L}.   
\eeq
The form of $U$ and $V$ in \eqref{E:UVhat} follows 
from Floquet's theorem and the observation that 
eigenfunctions are bounded, by definition.  
This decomposition has the benefit of admitting both $L$-periodic 
and $L$-anti-periodic eigenfunctions when $\mu=0$. Recall that $\psi$ 
is typically only quasiperiodic.  Allowing $\mu$ to be different from 0 
gives rise to solutions that are either quasiperiodic or have period 
greater than $2L$.  

Substitution of \eqref{E:phihat} and \eqref{E:UVhat} into \eqref{E:neval} 
and equating Fourier coefficients allows us to write equations for $U_n$ and $V_n$ as a coupled 
bi-infinite system of difference equations given by 

\begin{subequations}
\label{E:nevald}
\begin{multline}
\label{E:ndL1}
-\left(\lambda + \beta \rho^2 - \alpha\left(i\mu+ \frac{i n \pi}{L}\right)^2 \right) 
     U_n
+ 3 \gamma 
         \sum_{m = -\infty}^{\infty} Q_{\frac{n-m}{2}} U_m
- \alpha c^2  
         \sum_{m = -\infty}^{\infty} S_{\frac{n-m}{2}} U_m \\
 + 2 \alpha c 
         \sum_{m = -\infty}^{\infty} T_{\frac{n-m}{2}} V_m
- 2 \alpha c \left(i\mu + \frac{i n \pi}{L}\right) 
          \sum_{m = -\infty}^{\infty} R_{\frac{n-m}{2}} V_m 
    = 
        \Omega V_n, 
\end{multline}
\begin{multline}
\label{E:ndL0}
\left(\lambda + \beta \rho^2 - \alpha\left(i\mu+ \frac{i n \pi}{L}\right)^2 \right) 
        V_n
- \gamma
         \sum_{m = -\infty}^{\infty} Q_{\frac{n-m}{2}} V_m
+ \alpha c^2  
         \sum_{m = -\infty}^{\infty} S_{\frac{n-m}{2}} V_m \\
+ 2 \alpha c 
         \sum_{m = -\infty}^{\infty} T_{\frac{n-m}{2}} U_m
- 2 \alpha c \left(i\mu + \frac{i n \pi}{L}\right) 
          \sum_{m = -\infty}^{\infty} R_{\frac{n-m}{2}} U_m 
    = 
        \Omega U_n, 
\end{multline}
\end{subequations}
for all integers $n$.
Here  $\mu \in [\frac{-\pi}{K},\frac{\pi}{K})$ and
$Q_\frac{n-m}{2} = 0$ if $\frac{n-m}{2} \not\in \mathbb{Z}$, with 
$R_{(\cdot)},S_{(\cdot)}$ and $T_{(\cdot)}$ defined similarly.  
The system of equations \eqref{E:nevald} 
is {\em equivalent} to the original system \eqref{E:neval}. 

\subsection*{Remarks}
\begin{itemize}
\item 
 In practice, a pre-multiplication of the linear system by $\phi^4$ 
allows for the exact Fourier series expansion of $\phi^2, \phi^4$ and 
$\phi^6$ to be used.  This follows from the differential equations for 
$\sn(x,k)$ and Jacobi's series expansion of $\sn^2(x,k)$~\cite{jacobi}.
This pre-multiplication transforms the original 
eigenvalue problem into a generalized eigenvalue problem~\cite{GolubVanLoan}.
\item
  Note that Hill's methods enables one to compute the spectrum of a linear operator 
with periodic coefficients.  Despite the fact that the solution~\eqref{E:soln_form} is
typically quasiperiodic, the coefficient functions of the linear stability problem~\eqref{E:neval} 
are always $L/2$-periodic.   
\end{itemize}
 
\section{Numerical experiments} 
\label{s:experiments}

By choosing a finite number of Fourier modes, the 
exact bi-infinite system \eqref{E:nevald} is truncated. We explicitly 
construct and compute approximations to the spectral elements ({\em i.e.} eigenvalues or elements of the continous spectrum) of 
\eqref{E:neval} by finding the eigenvalues of the truncation of~\eqref{E:nevald}.  We consider all four cases individually: 
{\rm (I)} 
focusing in both $x$ and $y$
($\alpha = \beta =1$), 
{\rm (II)} 
focusing in $x$ and defocusing in $y$
($\alpha = -\beta =1$),
{\rm (III)}  
defocusing in $x$ and focusing in $y$ 
($-\alpha = \beta = 1$)  and finally,
{\rm (IV)} 
defocusing in both $x$ and $y$ 
($-\alpha = -\beta =1$). 

In each case, a large number of parameter values in the two-dimensional
parameter space shown in Fig.~\ref{F:parameter_space} 
was explored numerically. The ($k,B$)-parameter values considered correspond to NTP solutions, and do not include TP solutions or gray solitons. 
Approximately 5.2 million generalized eigenvalue problems were 
considered, the size of each determined by the cutoff
mode $N$ of the underlying Fourier series.  A truncation to $N$ positive 
Fourier modes reduces the exact bi-infinite system \eqref{E:nevald} to 
an $(4N+2)$-dimensional approximate problem. 
For several choices of $k$ and $B$, a value of $N=N(k,B)$ was chosen to ensure that the resulting eigenvalues had converged to within a measured tolerance. 
A simple polynomial was used to fit this data.
This information, and details related to other parameter ranges used in the 
experiments, are included in Table \ref{T:p_vals}.
In the table, $k$ is the elliptic modulus and $B$ is the offset parameter~(as in \eqref{E:defs}), 
$(4N+2)$ is the dimension used to 
approximate~\eqref{E:neval}, $\rho$ is the wavenumber of the perturbation in 
the $y$-dimension, and $\mu$ is the Floquet exponent. Lastly,
$\mathtt{linspace(a,b,m)}$ is a linearly spaced vector from $a$ to 
$b$ of length $m$, 
$\mathtt{logspace(a,b,m)}$ is a logarithmically spaced vector from $10^a$ to 
$10^b$ of length $m$ 
and {\tt ceil(x)} is the smallest integer greater than or equal to $x$.

\begin{table}[t]
\bc
\begin{tabular}{|c|c|l|}
\hline
Parameter & Description & \qquad Value \\
\hline
\hline
$k$ & Elliptic modulus & \, \tt{linspace(0,1,65)} \\ 
\hline
$B$ & Shift & \begin{tabular}{l}
              For $\alpha = -1:  -\mathtt{logspace(-8,0,65)} $ \\ 
              For $\alpha = \phantom{-}1: (2k^2 + \mathtt{logspace(-8,0,65)}) \cap (2k^2 , 2)$
              \end{tabular} \\ 
 
\hline
$N$ & Fourier cutoff & \begin{tabular}{l}
                        For $\alpha=-1: 15 + \mathtt{ceil(5 k^{5})}$ \\ 
                        For $\alpha= \phantom{-}1: 10 + \mathtt{ceil(25 k^{10})}$ 
                        \end{tabular} \\
\hline
$\rho$ & Perturbation wavenumber & \, \tt{linspace(0,4,65)} \\
\hline
$\mu$ & Floquet parameter & \, $\mathtt{linspace(-\frac{\pi}{K},\frac{\pi}{K},21)}$  \\
\hline
\end{tabular}
\ec
\caption{Parameter values and ranges used in numerical experiments. Only perturbations of NTP solutions are considered.{\label{T:p_vals}}}
\end{table}

\section{Observations from the numerical investigation}
\label{s:observations}

First and foremost, it should be stated that {\em none} of the NTP solutions 
considered here were found to be spectrally stable under one-dimensional ($\rho = 0$) or two-dimensional ($\rho \neq 0$) perturbations.  This establishes, at 
least numerically, that {\em all} one-dimensional traveling-wave solutions of NLS of the form~\eqref{E:soln_form} are spectrally
unstable with respect to either one- or two-dimensional perturbations.  At this point, it remains to investigate the nature of the 
instabilities and their corresponding growth rates, so as to better understand the dynamics of this important 
class of solutions of the NLS equation.

Using Hill's method we numerically considered the 
instabilities due to
perturbations with wavenumber $\rho \in [0,4]$. Note that $\rho=0$ 
corresponds to one-dimensional perturbations.
For each NLS equation ({\em i.e.} for each choice of $\alpha,\beta = \pm 1$), 
and solution ({\em i.e.} each parameter pair $(k,B)$), and for each perturbation of wavenumber $\rho$, an equally-spaced 
sequence of Floquet parameters 
$\mu$ was chosen from the interval $[-\frac{\pi}{K},\frac{\pi}{K}]$. 
The generalized  eigenvalues and eigenvectors were computed for the 
matrix that results from a truncation of~\eqref{E:nevald}.  The 
generalized eigenvalues are approximations of spectral elements 
of \eqref{E:neval}, and an approximation of the corresponding 
eigenfunctions may be reconstructed from the eigenvectors.

Since a single eigenvalue with positive real part leads to 
instability of the system, the approximate eigenvalue with largest 
real part over 
all choices of $\mu$ was recorded for each $(k,B,\rho)$ triplet. That is,
we compute
\begin{equation}
\Omega_{\mbox{\scriptsize growth}}(k,B,\rho) = 
\max_{\mu \in [-\pi/K,\pi/K]} \, \mathrm{Re} \left[ \Omega(k,B,\rho,\mu) \right] ,
\end{equation}
which we call the (most unstable) growth rate.
which we call the (most unstable) growth rate.
The value $\Omega_{\mbox{\scriptsize growth}}$ represents the largest exponential growth rate a given NTP solution with 
parameters $(k,B)$ will experience when perturbed with transverse wavenumber $\rho$.
It also allows us to determine
the perturbation to which the NTP solution is spectrally the most unstable.    
We reduce the dimension further by computing the  
largest such growth rate over all sampled perturbation wavenumbers $\rho$.
This quantity, 
\begin{equation}
\Omega_{\max}(k,B) = \max_{\rho \in [0,4]}\Omega_{\mbox{\scriptsize growth}}(k,B,\rho),
\end{equation}
the maximal growth rate over all $\rho$, is plotted in the first columns of 
Figs.~\ref{F:focus_growth} and~\ref{F:defocus_growth}.  The value 
$\Omega_{\max}$ represents the maximal exponential growth 
rate that a solution with parameters $(k,B)$ can undergo in the range examined.
We also recorded the minimum growth rate over all $\rho$, 
\begin{equation}
\Omega_{\min}(k,B) = \min_{\rho \in [0,4]}\Omega_{\mbox{\scriptsize growth}}(k,B,\rho),
\end{equation}
to verify that all solutions are unstable with respect to every sampled perturbation.

Every point plotted in Figs.~\ref{F:focus_growth} and \ref{F:defocus_growth} corresponds 
to an NLS solution for which we considered the linear stability analysis, and the boundaries in the figures are the boundaries of 
the regions represented in Fig.~\ref{F:parameter_space}, corresponding to 
limiting TP solutions.
Fig.~\ref{F:focus_growth} corresponds to the $x$-focusing ($\alpha = 1$)
parameter range $(k,B) = (0,1) \times (2k^2,2)$ in the $\alpha = 1$ case 
of Fig.~\ref{F:parameter_space}a. 
The one-to-one transform $T_f(B) = (B-2k^2)/(2-2k^2)$ is used to normalize the 
range of $B$.  This maps the interval $[2k^2,2]$ to $[0,1]$.
Figure~\ref{F:defocus_growth} corresponds to the $x$-defocusing ($\alpha = -1$)
parameter range of $(k,B) = (0,1) \times (-1,0)$ shown in 
Fig.~\ref{F:parameter_space}b.  The 
transform $T_d(B) = -B$ is used in Fig.~\ref{F:defocus_growth}.
A log$_{10}$ scale is used in the vertical dimension of 
Figs.~\ref{F:focus_growth} and~\ref{F:defocus_growth}.
This causes the panels of Fig.~\ref{F:focus_growth} to become increasing sparse in their lower right corners.  
The right-hand panels of Figs. \ref{F:focus_growth} and \ref{F:defocus_growth} indicate the wavenumber $\rho$ that leads to 
maximal growth shown in the left-hand panels.  
Recall that our computations were limited to $\rho \in [0,4].$

\subsection{Case {\rm I}: Elliptic setting with $\alpha = \beta = 1$}
\label{ss:case1}

Panels {\rm I}a and {\rm I}b of Fig.~\ref{F:focus_growth} summarize 
some properties of the 
computed instabilities in the case of focusing in both the $x$- and 
$y$-dimensions.   
The lower boundary of the plot corresponds to $B = 2k^2 + (10^{-8})$, and is 
therefore only slightly away (in the parameter space of $B$) from a cn-type solution.  The upper boundary is close to 
dn-type solutions, with $B=1.99$.
The left boundary of the plots, 
where $k = 0.01$, represents a region in parameter space near to Stokes' 
wave solutions. The entire right-hand boundary, where $k=0.99$, is 
near to the bright soliton limit case which occurs at $(k,B)=(1,2)$.  

A distinct ridge of large instability is noticeable in the plot 
of $\Omega_{\max}$ in panel {\rm I}a of Fig.~\ref{F:focus_growth}.  The ridge appears to begin
near the zero solution at $(k,B) = (0,0)$, and remains close to the cn 
limit boundary (within approximately .02 units, remembering the $\log_{10}$ scaling) as $k$ increases. 
Moving away from the cn boundary results in the rapid increase of $\Omega_{\max}$.  Movement away from the dn boundary results in a slower increase in the value of $\Omega_{\max}$, as does moving away from the Stokes' wave boundary for $B$ greater than approximately 0.001.
The maximum value of $\Omega_{\max}$ over the sampled $(k,B)$ space, given by 
$R_{\max}=5.666$, is reached at $(k,B) = (0.99,1.98)$. This growth rate should be compared to the maximal growth rate of the corresponding TP case~\cite{JCBDtp} which is $R_{\max}= 1$.  The minimum 
($R_{\min} = 0.015693$) occurs for $(k,B) = (0.01,0.01)$ for $\rho = 4$. 
Note that this is on the boundary of the computational domain and that at 
$(k,B) = (0,0)$, corresponding to the spectrally stable $\psi \equiv 0$ 
solution, $R_{\min} = 0.$

In panel {\rm I}b, the wavelength corresponding to the maximal growth of  
{\rm I}a is given.  In this case, the maximum instability occurs 
for the shortest wavelength samples, $\rho = 4$.
This indicates that there is a strong short-wavelength instability.

\subsection{Case {\rm II}: Hyperbolic setting with $\alpha = -\beta = 1$}
\label{ss:case2}

\begin{figure}[t]
\begin{center}
\psfrag{ab 1 1}{\,}
\psfrag{ab 1 -1}{\,}
\psfrag{Wave number 1 1}{\,}
\psfrag{Wave number 1 -1}{\,}
\psfrag{k}{\,}
\psfrag{flog}{\hspace*{-1cm}$\log_{10} \,T_f(B)$}
\begin{tabular}{ccc}
Ia. $\Omega_{\max}$ with $(\alpha,\beta) =  (1,+1)$ & \, &
Ib. Corresponding $\rho$, $(\alpha,\beta) =  (1,+1)$  \\
 \epsfig{file=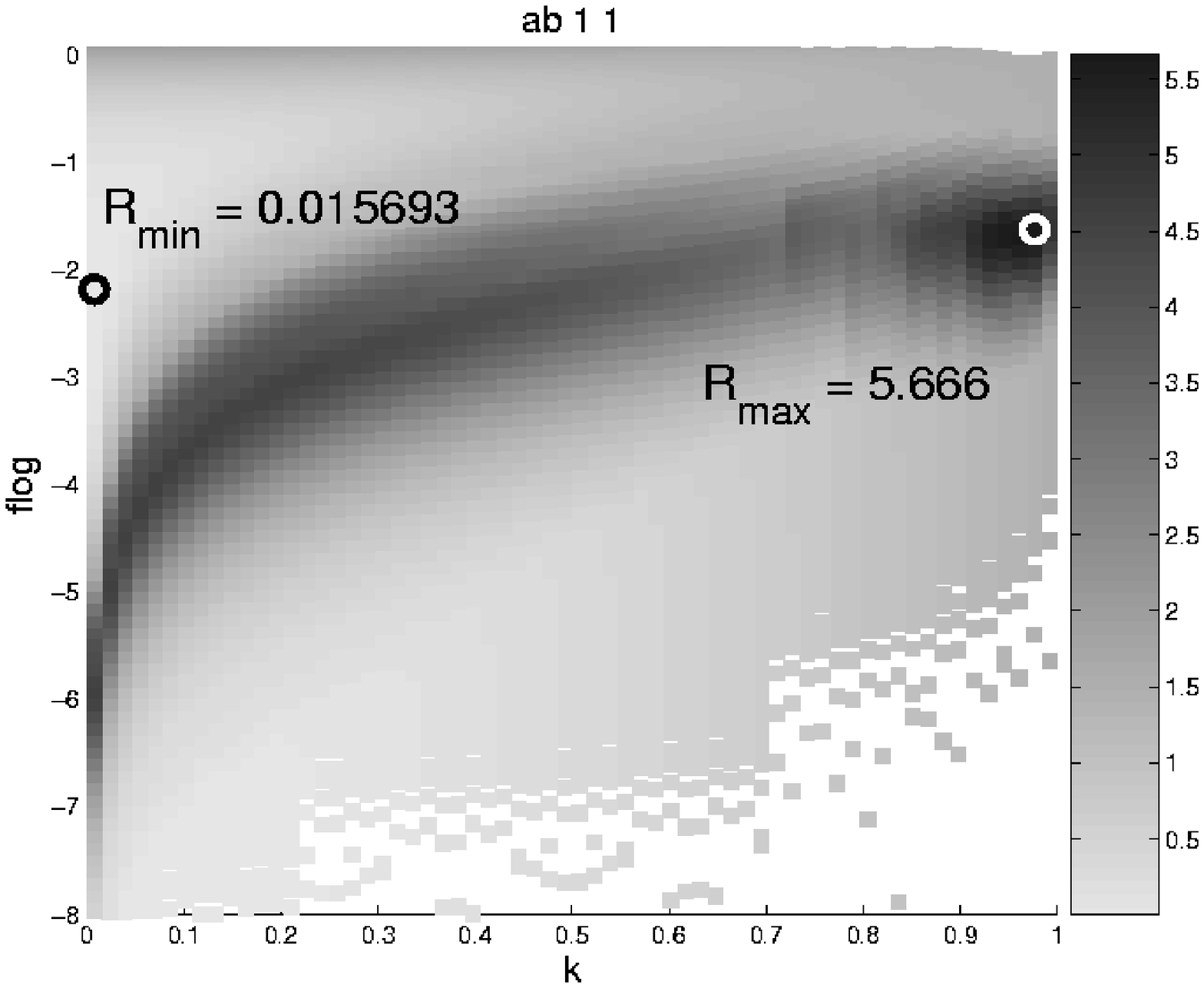, width = 6cm} & &
 \epsfig{file=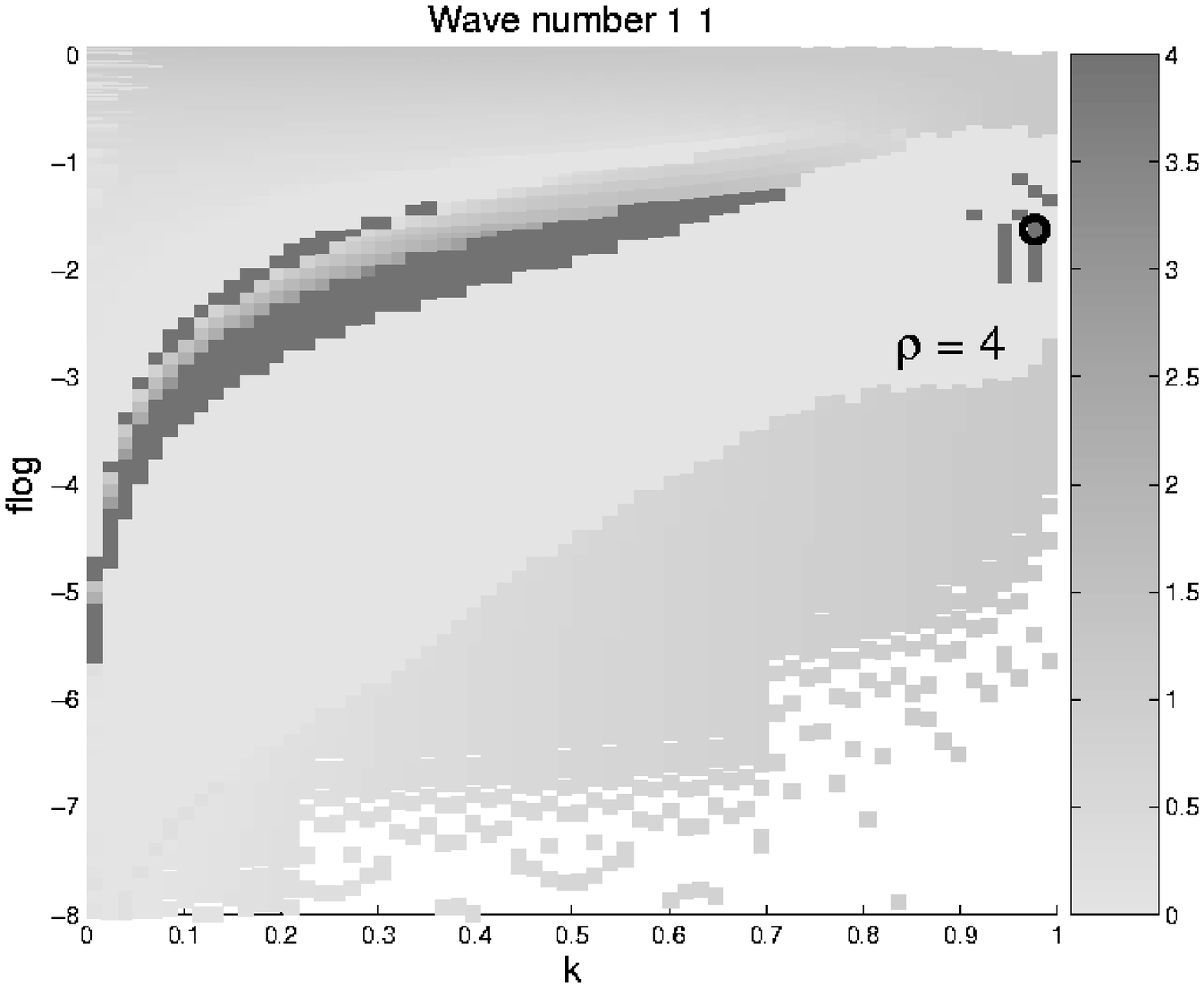, width = 6cm}  \\
k & & k \\
IIa. $\Omega_{\max}$ with $(\alpha,\beta) =  (1,-1)$ & & 
IIb. Corresponding $\rho$, $(\alpha,\beta) =  (1,-1)$ \\ 
 \epsfig{file=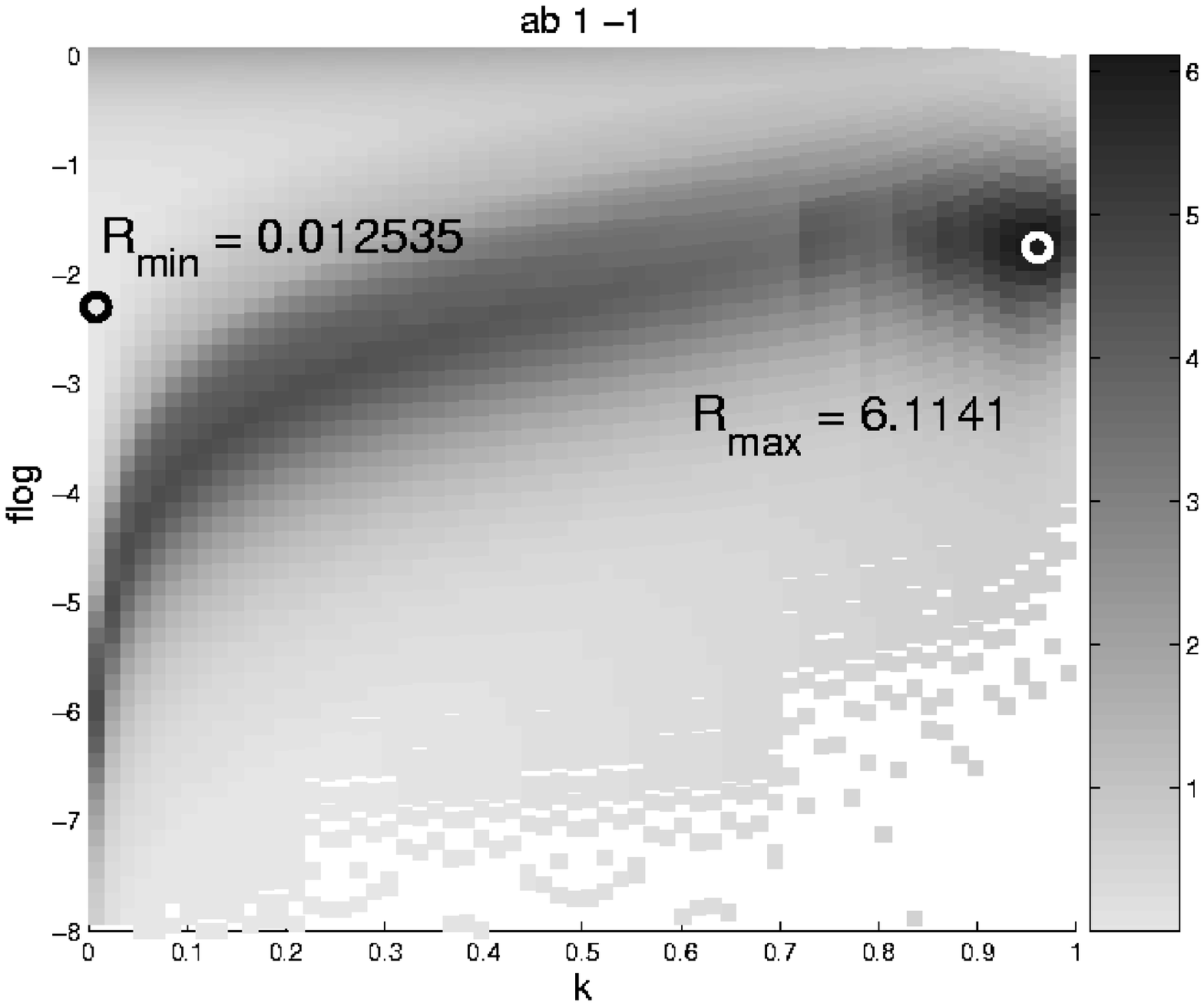, width = 6cm}  & &
 \epsfig{file=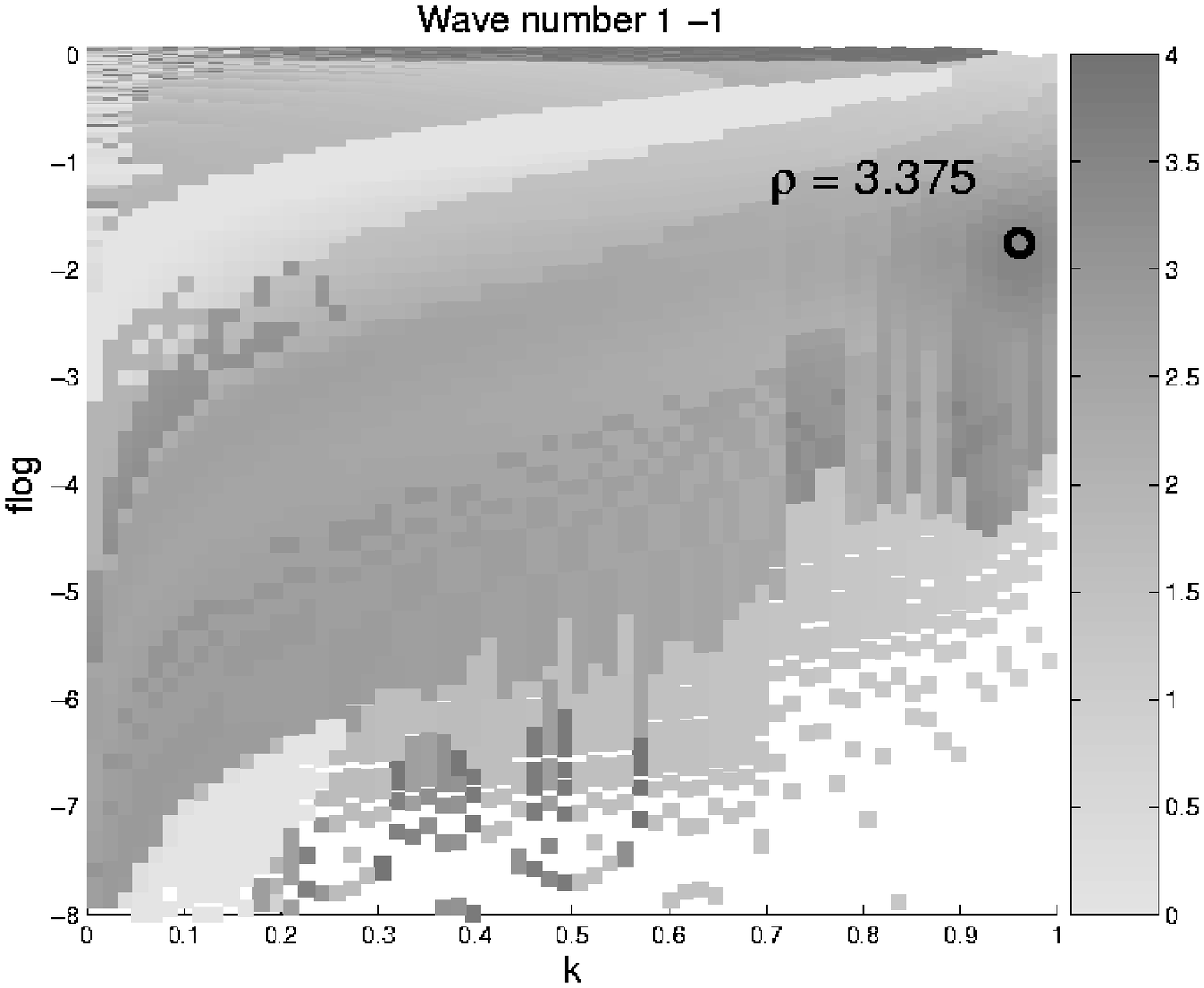, width = 6cm}  \\
k & & k
\end{tabular} 
\caption{Focusing in the $x$-dimension.  The first column contains 
surface plots of 
$\Omega_{\max}$ vs. $(k,B)$ and  
the second column contains the surface plots of the maximizing wavenumber $\rho$ vs. $(k,B)$. 
R$_{\max} = \max_{k,B} \Omega_{\max}$  and 
R$_{\min} = \min_{k,B} \Omega_{\min}$. 
White space corresponds to ($k,B$)-parameter space that was {\em not} sampled.{\label{F:focus_growth}}}
\end{center}
\end{figure}

Panels {\rm II}a and {\rm II}b of Fig. \ref{F:focus_growth} summarize 
some properties of 
the computed instabilities in the case of focusing in the $x$-dimension 
and defocusing in the $y$-dimension.
The lower boundary of the plot corresponds to $B = 2k^2 + (10^{-8})$, and is 
therefore only slightly away (in the parameter space of $B$) from a cn-type solution.  The upper boundary is close to 
dn-type solutions, with $B=1.99$.
The left boundary of the plots, 
where $k = 0.01$, represents a region in parameter space near to Stokes' 
wave solutions. The entire right-hand boundary, where $k=0.99$, is 
near to the bright soliton limit case which occurs at $(k,B)=(1,2)$.  

As in Case {\rm} I, a ridge of large growth rate is noticeable in 
the growth plot shown in panel {\rm II}a.  The ridge appears to begin
near the zero solution at $(k,B)=(0,0)$, and remains close to the cn-type 
limit boundary (within approximately .02 units, remembering the 
$\log_{10}$ scaling) as $k$ increases.  This ridge has a local minimum 
near $k=0.7$ and increases to a global (over all admissible 
$(k,B)$-parameter space) maximum at $k = 0.96$.
As in the setting above, moving away from the cn-type boundary results 
in a rapid increase of $\Omega_{\max}$.  Moving away from the dn 
boundary results in a slower increase in the value of $\Omega_{\max}$. 
For $B>0.001$, moving away from the boundary result in a similar slow 
increase in $\Omega_{\max}$.  
For $k > 0.96$, it appears that the limiting value of $\Omega_{\max}$ is 
consistent with the bright soliton results of~\cite{kivpel,JCBDsoliton}.
The maximum ($R_{\max} = 6.1141$) and minimum ($R_{\min} = 0.012535$) 
growth rates span a slightly larger range than the similar values 
in Fig.~Ia. 
These occur at $(k,B) = (0.01,0.01)$ and
$(k,B) = (0.96,1.98)$, respectively.
The maximal growth rate should be compared to the maximal growth rate 
of the corresponding TP case~\cite{JCBDtp} which is $R_{\max}= 1$. 

In panel {\rm II}b, the wavelength corresponding to the maximal 
growth $R_{\max}$ of {\rm II}a is given.  In this case, the maximum 
instability of $R_{max}$ occurs for $\rho = 3.375.$  The surface shown in 
{\rm II}b appears to be more smooth than the surface of {\rm I}b. 

\subsection{Case {\rm III}: Hyperbolic setting with $-\alpha = \beta = 1$}
\label{ss:case3}

Panels {\rm III}a and {\rm III}b of Fig. \ref{F:defocus_growth} summarize 
some properties of 
the computed instabilities in the case of defocusing in the $x$-dimension 
and focusing in the $y$-dimension.
The lower limit of the plot corresponds to $B = - (10^{-8})$, and so is 
just slightly away from the sn-type solution. The left boundary of the plots, 
where $k = 0.01$, represents a region in parameter space near Stokes' wave solutions, while $k=0.99$ on the right boundary is near to the gray soliton limit.

A distinct ridge of large instability is noticeable in the growth plot displayed in panel 
{\rm III}a.  The ridge appears to begin
near the zero solution limit at $(k,B) = (0,0)$, and remains close to the sn 
limit boundary (within approximately .02 units, remembering the $\log_{10}$ scaling) as $k$ increases. It quickly reaches 
the global maximum (over all admissible $(k,B)$-parameter space) of $R_{\max} = 7.6375$ at $k = 0.02$ and $B=-0.0001$.
 The ridge then  
decreases in amplitude as $k$ increases towards 1.
Moving away from the sn-type boundary results in a rapid increase of $\Omega_{\max}$.  Moving away from the dn-type boundary results in a slower increase in the value of $\Omega_{\max}$. Similarly, the increase is slower when moving away from the Stokes' wave limit for $B<-0.001$.
The maximum exponential growth rate, $R_{\max} = 7.6375$, occurs for $(k,B) = (0.02,-0.00001)$.  
This growth rate should be compared to the maximal growth rate 
of the corresponding TP case~\cite{DS} which is $R_{\max}= 1$. 
The minimum exponential growth, $R_{\min}=0.015578$ is found at $(k,B) = (0.01,-0.9)$.  Both the maximum and minimum are located near the  
Stokes' wave boundary. 
By restricting $\rho=0$ and allowing $B$ to approach zero, $\Omega_{\max} \to 0$, and the one-dimensional stability result of the sn-type TP solution of~\cite{JCBDtp} is recovered.

The plot {\rm III}b indicates short-wave perturbations lead to large values of 
$\Omega_{\max}$. The largest growth occurs for a 
perturbation with wavenumber of $\rho = 3.625$. 

\subsection{Case {\rm IV}: Elliptic setting with $-\alpha = -\beta = 1$}
\label{ss:case4}

\begin{figure}
\begin{center}
\psfrag{ab -1 1}{\,}
\psfrag{ab -1 -1}{\,}
\psfrag{Wave number -1 1}{\,}
\psfrag{Wave number -1 -1}{\,}
\psfrag{k}{\,}
\psfrag{dlog}{\hspace*{-1cm}$\log_{10}(-B)$}
\begin{tabular}{ccc}
IIIa. $\Omega_{\max}$ with $(\alpha,\beta) =  (-1,+1)$ &  \, & 
IIIb. Corresponding $\rho$, $(\alpha,\beta) =  (-1,+1)$ \\ 
\epsfig{file=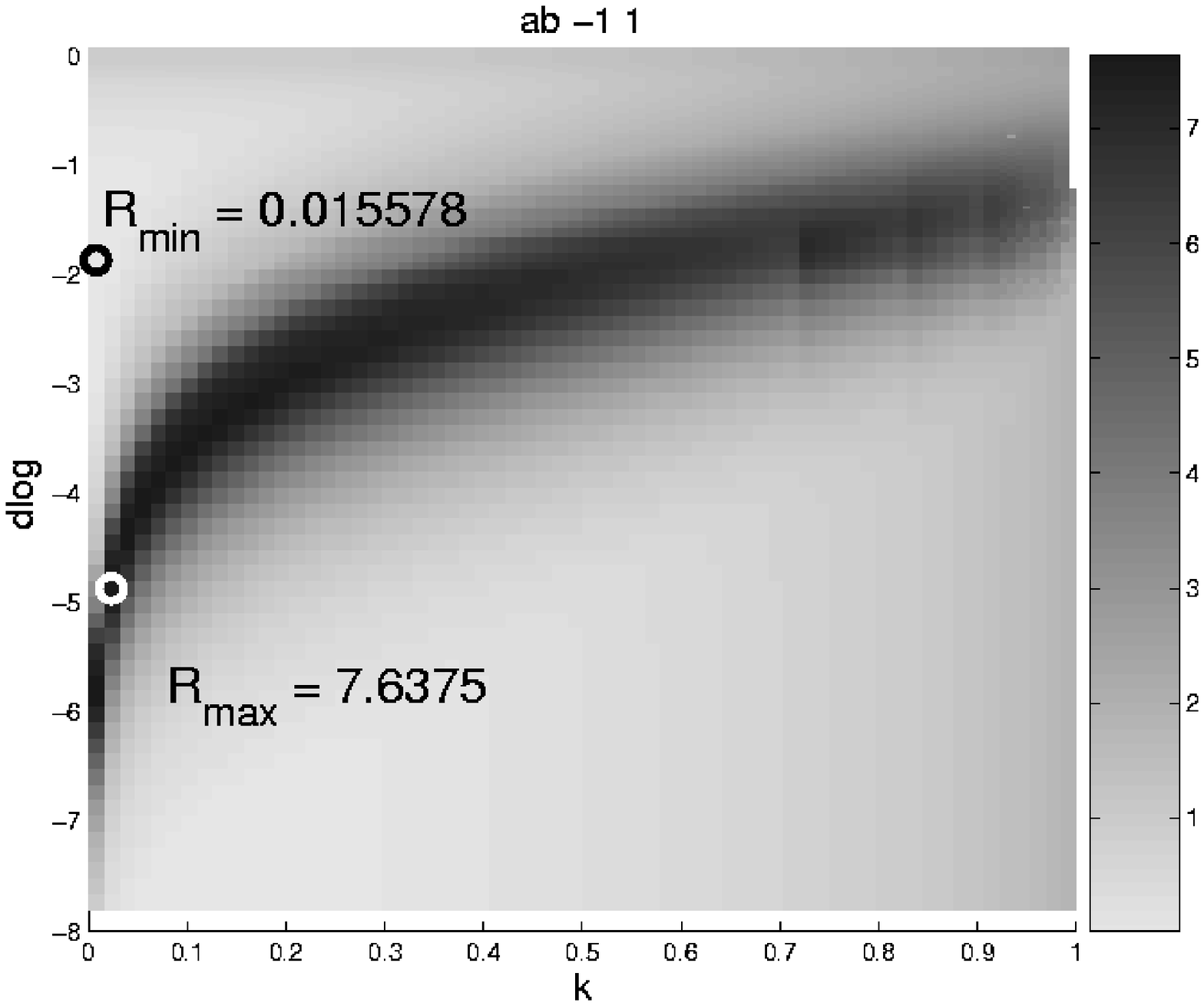, width = 6cm}  & & 
\epsfig{file=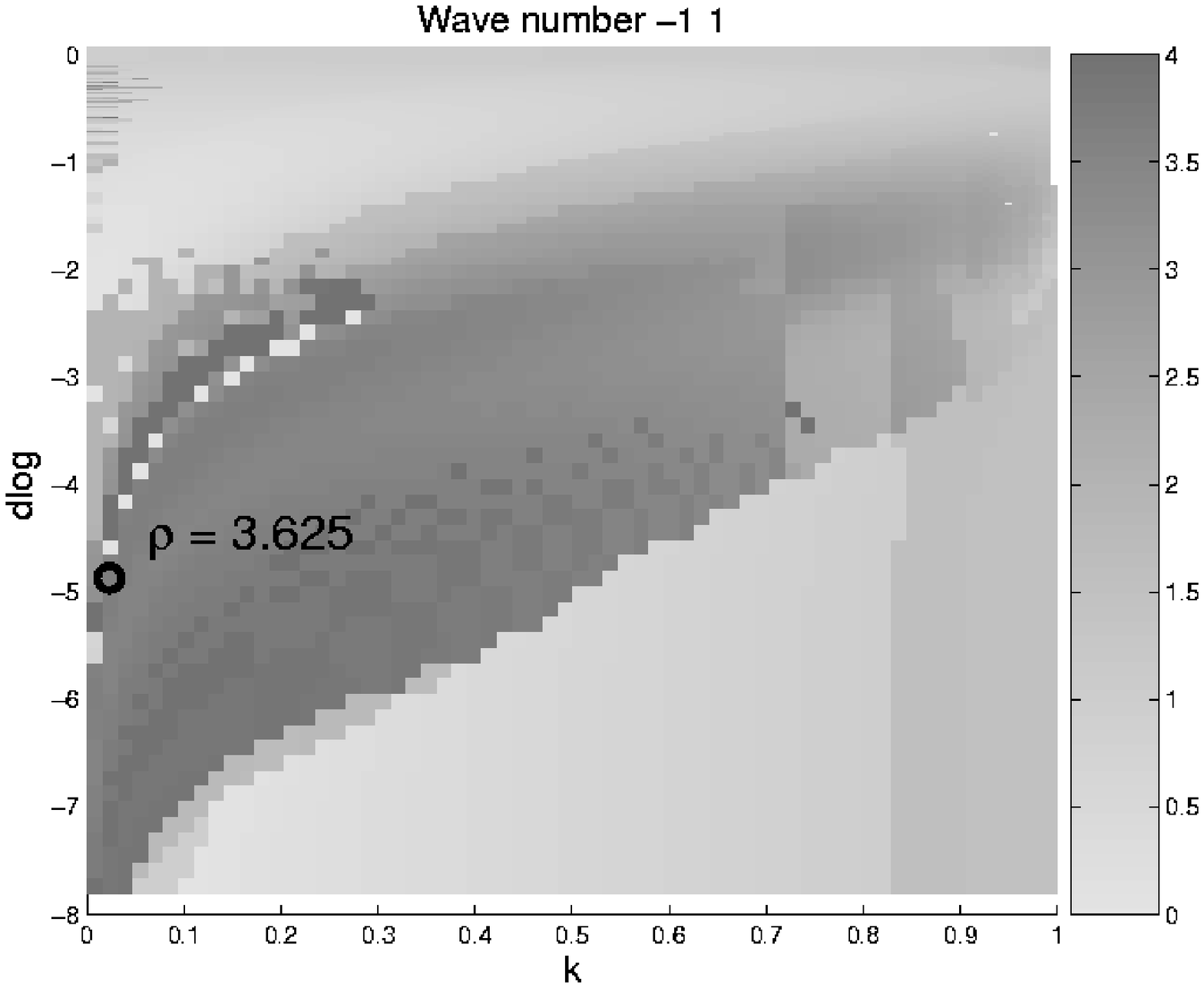, width = 6cm} \\
k & &  k \\
IVa. $\Omega_{\max}$ with $(\alpha,\beta) =  (-1,-1)$ &  &
IVb. Corresponding $\rho$, $(\alpha,\beta) =  (-1,-1)$ \\ 
\epsfig{file=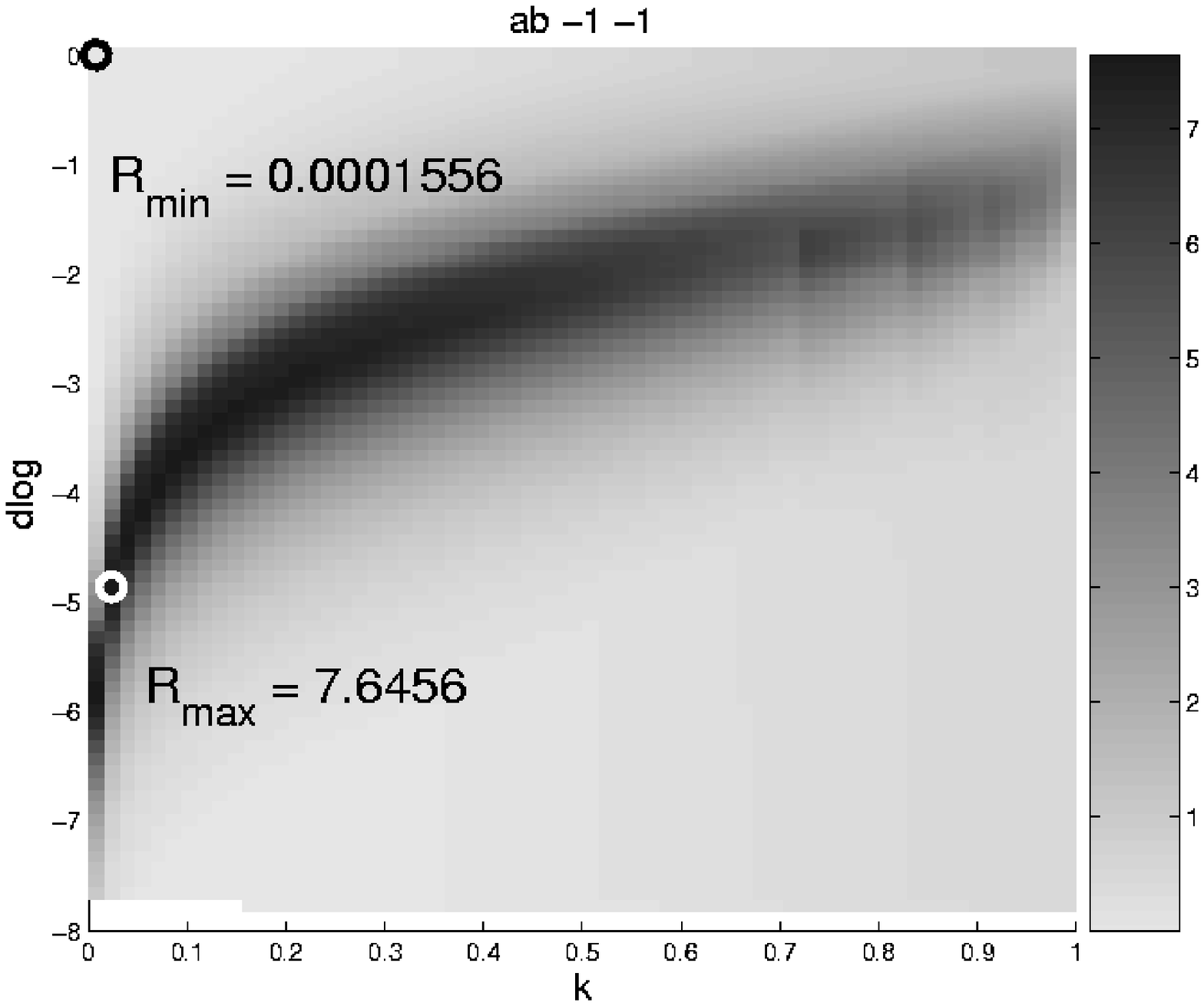, width =  6cm}  & &
 \epsfig{file=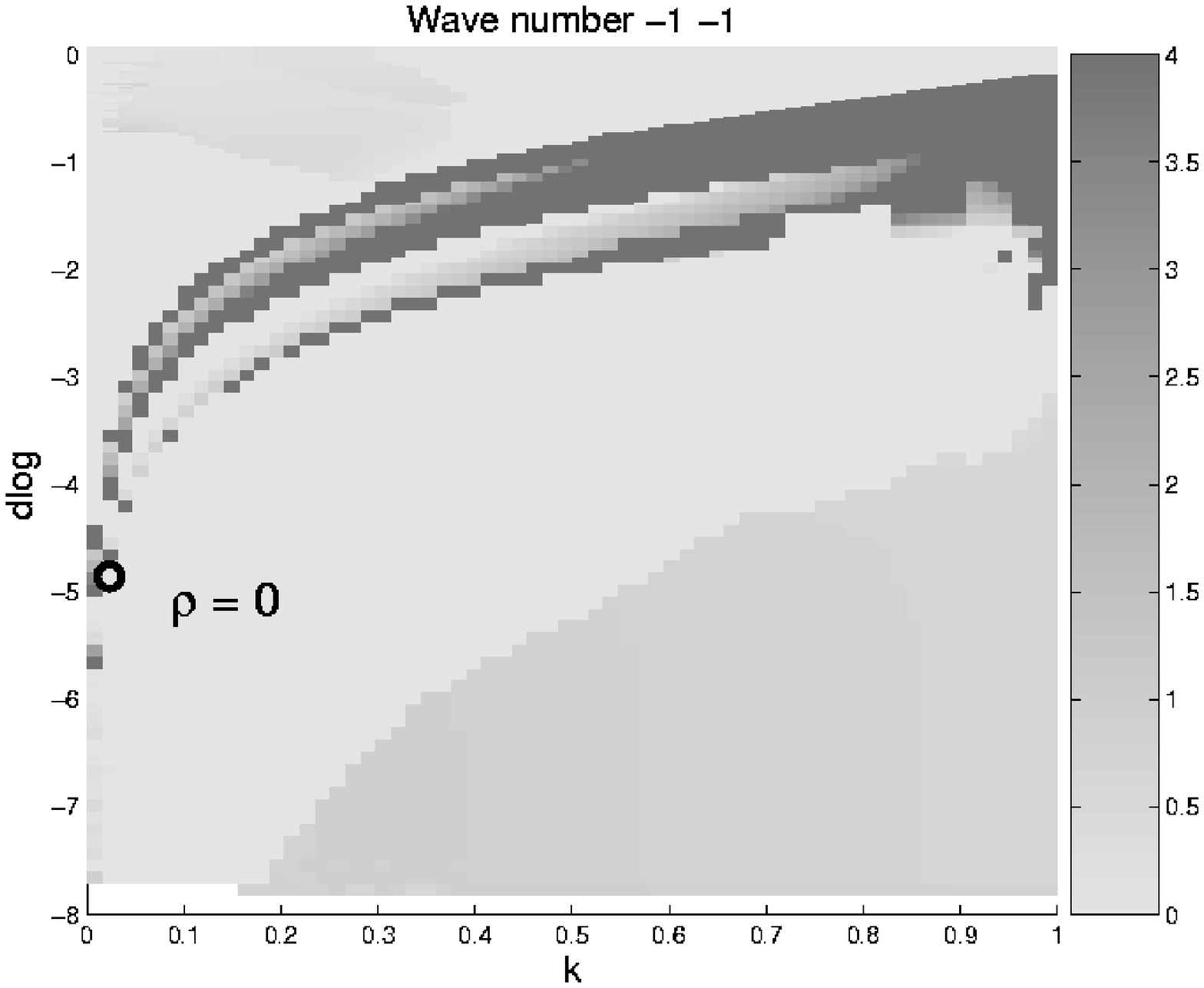, width = 6cm}\\
k & & k
\end{tabular} 
\caption{Defocusing in the $x$-dimension. The first column contains surface plots of 
$\Omega_{\max}$ vs. $(k,B)$ and the second column 
contains surface plots of the maximizing wavenumber $\rho$ vs. $(k,B)$. 
R$_{\max} = \max_{k,B} \Omega_{\max}$  and 
R$_{\min} = \min_{k,B} \Omega_{\min}$. 
{\label{F:defocus_growth}}}
\end{center}
\end{figure}

Panels {\rm IV}a and {\rm IV}b of Fig.~\ref{F:defocus_growth} 
summarize some properties of the computed instabilities in the case of defocusing 
in both the $x$- and $y$-dimensions.
The lower limit of the plot corresponds to $B = - 10^{-8}$, and so is 
just slightly away from the sn-type solution. The left boundary of the plots, 
where $k = 0.01$, represents a region in parameter space near to Stokes' wave solutions, while $k=0.99$ on the right boundary is near to the gray soliton limit.

A distinct ridge of large instability is noticeable in the growth plot 
shown in panel {\rm IV}a.  The ridge appears to begin
near the trivial limit $k=0$ and $B=0$, and remains close to the sn 
limit boundary (within approximately .02 units, remembering the $\log_{10}$ scaling) as $k$ increases, to reach 
a global maximum at $k = 0.02$ and $B=-0.0001$. The ridge then appears to 
decrease in amplitude as $k$ increases towards 1.
As in Case III, moving away from the sn-type boundary results in a rapid increase of $\Omega_{\max}$.  Moving away from the dn-type boundary results in a slower increase in the value of $\Omega_{\max}$. The same is true when moving away from the Stokes' boundary, when $B$ is less than approximately -0.001.
The maximum exponential growth rate, $R_{\max} = 7.6456$, and the 
minimum, $R_{\min}=0.0001556$, span a slightly larger range of values than do the values of $\Omega_{\max}$ in panel IIIa. The maximum and minimum values are obtained at $(k,B) = (0.01,-0.00009)$ and $(k,B) = (0.01,-1)$, respectively.  Both are located near the Stokes' wave boundary.  
The maximal growth rate should be compared to the maximal growth rate 
of the corresponding TP case~\cite{DS} which is $R_{\max} \approx 0.26 $. 
As in Case III, restricting $\rho=0$ and allowing $B$ to approach zero results in $\Omega_{\max} \to 0$, and the one-dimensional stability result of~\cite{JCBDtp} for the sn-type TP solution is recovered.    

In panel {\rm IV}b, wavenumbers corresponding to 
$\Omega_{\max}$ of {\rm IV}a are given.  It appears that a majority of 
the large values of $\Omega_{\max}$ are  
attributable to small-$\rho$ (long-wave) perturbations. In fact,
the largest growth occurs for $\rho = 0$, the one-dimensional perturbation. 
This should be contrasted with the three previous cases, where short-wavelength two-dimensional perturbations with wavenumber $\rho>3$ were associated with the largest $\Omega_{\max}$ values.

\section{Summary}
\label{s:summary}

In this paper, we considered the spectral instability of one-dimensional 
traveling-wave nontrivial-phase (NTP) solutions of the cubic nonlinear Schr\"{o}dinger equation.
Such solutions are expressed in terms of Jacobi elliptic functions.
An exact spectral form of the linearized operator is 
truncated and used to construct an associated generalized eigenvalue problem. 
The positive real parts of the resulting eigenvalues were used to
determine that there are {\em no} stable NTP solutions.

Numerical results indicate a well-defined ridge of large growth rate 
located in the $(k,B)$-parameter region associated 
with nontrivial-phase solutions. This implies that 
the most unstable NTP solutions are more unstable than any TP solution,
in the sense that they exhibit larger exponential growth rates.
Further, for all cases the 
exponential growth rate $\Omega_{\max}$ increases when 
moving away from the limiting TP solutions.  This divergence is gradual in some cases, 
but very sharp in other cases, as discussed above.

In summary, numerical evidence suggests that all bounded, 
nontrivial-phase one-dimensional traveling-wave solutions of the 
cubic NLS equation are unstable with respect both one-dimensional 
and two-dimensional perturbations.

\subsection*{Acknowledgements}
We gratefully acknowledge support from the National Science Foundation: 
NSF-DMS-FRG:0139847 (RJT), NSF-DMS-FRG:0139771 (JDC) and NSF-DMS-FRG:0351466 (BD).

\bibliographystyle{plain}
\bibliography{ntppaper}

\end{document}